\documentstyle[preprint,prl,aps]{revtex}
\tighten
\begin{document}
\draft
%\preprint{LPTENS 95/??}
\title{Non Fermi Liquid Behaviour near a $T=0$ spin-glass transition}
\author{Anirvan M. Sengupta}
\address{AT\&T Bell Laboratories\\
600 Mountain Avenue, Murray Hill, NJ 07974; USA.\\
e-mail: anirvan@physics.att.com}
\author{Antoine Georges}
\address{ CNRS-Laboratoire de Physique Th\'{e}orique de l'Ecole
Normale Sup\'{e}rieure$^1$
\\
24, rue Lhomond; 75231 Paris Cedex 05; France
\\
e-mail: georges@physique.ens.fr}
\date{April 27, 1995}
\maketitle
\begin{abstract}
In this paper we study the competition between the Kondo effect and RKKY
interactions near the zero-temperature quantum critical point of an
Ising-like metallic spin-glass. We consider the mean-field behaviour
of various physical quantities. In the `quantum- critical regime'
non-analytic corrections to the Fermi liquid behaviour are found for
the specific heat and uniform static susceptibility, while the resistivity
and NMR relaxation rate have a non-Fermi liquid dependence on temperature.
\vspace{3cm}

\begin{description}
\item[$^1$]  Unit\'{e} propre du CNRS (UP 701) associ\'{e}e \`{a} l'ENS
et \`{a} l'Universit\'{e} Paris-Sud
\end{description}

\end{abstract}

% \pacs{ PACS numbers:}

\narrowtext
\newpage

\section{Introduction}

The interplay between Kondo screening of localised spins by conduction
electrons, and ordering of these spins due to the RKKY interaction is
a central issue in heavy fermion physics. Recently, a class of systems
has been studied \cite{MBM,VL} in which the ordering temperature is
driven to zero as a function of concentration, and the paramagnetic metal
displays non Fermi liquid (NFL) behaviour near this $T=0$ quantum critical
point. The compound $Y_{1-x}U_xPd_3$ is one of the best documented among
these systems \cite{MBM,SML}. In this case, the low-temperature
ordered state is reported to be a spin-glass for $x>x_c\simeq 0.2$ ,
while the system remains a paramagnet down to the lowest temperature
studied for $x<x_c$. It is still debated \cite{DC,AT}
whether the NFL behaviour of the $Y_{0.8}U_{0.2}Pd_3$ system is a single-ion
effect or results from the above competition and the proximity
of the $T=0$ critical point. The aim of this paper is not to resolve this
debate for this particular system, but to demonstrate that NFL behaviour
is indeed a generic feature of the vicinity of a $T=0$ paramagnetic metal to
metallic spin-glass transition. This will be shown by solving specific
models at mean-field level.

\section{Models}

The models that we shall study are mean-field versions of the Kondo
lattice, with an additional quenched randomness on the exchange
interactions between localised spins.
We consider localised spins $\vec{S}_i$  on a fully connected lattice of
$N$ sites $i=1,\cdots,N$. These spins interact with a bath of conduction
electrons. In the model that we shall consider first, a major simplification
will be made: the conduction electron bath will be
assumed to consist of {\it independent} "reservoirs" of
electrons, with no communication between the reservoirs at different sites.
The effect of releasing this simplifying assumption will be discussed
in Sec.\ref{mod2} at the end of this paper.
The conduction electrons will be denoted by $c^{(i)}_{k\alpha}$,
where $\alpha=\uparrow,\downarrow$ is a spin
index, $k$ labels the conduction band orbitals, and the site index $(i)$
labels the reservoir associated with site $i$. The hamiltonian of the
model reads:
\begin{equation}
H_1\,=\,\sum_{k,\alpha,i}\epsilon_k c^{(i)+}_{k\alpha}c^{(i)}_{k\alpha} +
J_K \sum_i \vec{S}_i\cdot\vec{s}(i) -
\sum_{i<j}J_{ij}S_i^z\cdot S_j^z
\label{ham1}
\end{equation}
In this expression,
$\vec{s}(i)\equiv\sum_{\alpha\beta}\sum_{k,k'}{{1}\over{2}}c^{(i)+}_{k\alpha}
\vec{\sigma}_{\alpha\beta}c^{(i)}_{k'\beta}$ is the conduction electron
spin-density at site $i$, and $J_K$ the strength of the Kondo coupling
between the localised spins and the conduction electrons (taken to be
antiferromagnetic). Besides this coupling, the spins have a direct
interaction between one another: the $J_{ij}$'s are quenched independent
random variables with a distribution $P(J_{ij})\sim\exp(-J_{ij}^2/4NJ^2)$.
A further simplification of our model is that only the Ising part of the
exchange interaction has been included. For $J_K=0$, the model reduces to
the Sherrington-Kirkpatrick model of a classical Ising spin-glass \cite{SK}
with a freezing transition at $T_c=J$. In contrast, for $J=0$, we have
a system of independent localised spins, each one being quenched by the
Kondo effect with its conduction electron reservoir, and the system has
no long-range order down to $T=0$. We are interested in the intermediate
behaviour where the spin-glass freezing due to the random exchange competes
with the local Kondo effect.

Because the model is fully connected, it can be reduced to a single-site
problem after taking the (quenched) average over the realizations of
the random couplings. In order to describe
also the spin-glass phase, we may introduce replicated variables labeled
by indices $a,b=1,\cdots,n$. Using standard techniques \cite{BM,GK,GKS},
the single-site effective action is found to be (with obvious notations):
\begin{eqnarray}
& S_{eff} = -\int_0^{\beta} d\tau \int_0^{\beta} d\tau' \sum_{\alpha,a}
c_{\alpha,a}^+(\tau) {\cal G}_0^{-1} (\tau-\tau') c_{\alpha,a}(\tau')
+ J_K \int_0^{\beta} d\tau\sum_a \vec{S_a}(\tau)\cdot\vec{s_a}(\tau)
\nonumber\\
& - J^2 \int_0^{\beta} d\tau \int_0^{\beta} d\tau'\sum_{a,b}
S_a^z(\tau) D_{ab}(\tau-\tau') S_b^z(\tau')
\label{seff}
\end{eqnarray}
In this equation, ${\cal G}_0$ is simply the bare on-site propagator
for the conduction electrons:
${\cal G}_0\equiv \sum_k 1/(i\omega_n-\epsilon_k)$, which can be taken
to have the characteristic form associated with a flat band without loss
of generality: ${\cal G}_0(i\omega_n)^{-1}=i\Gamma \mbox{sgn}(\omega_n)$.
The non-local spin interaction $D_{ab}$ must satisfy the
following self-consistency
condition in the thermodynamic limit $N\rightarrow\infty$:
\begin{equation}
D_{ab}(\tau-\tau') = < T S^z_a(\tau) S^z_b(\tau') >_{S_{eff}}
\label{scspin}
\end{equation}
In the following, we shall mainly be concerned with the paramagnetic phase
in which replica symmetry holds, so that $D_{ab}(\tau)$ is non-zero only for
$a=b$. In the spin-glass phase, $D_{ab}$ is non-zero (but $\tau$-independent)
for $a\neq b$, and the Edwards-Anderson order parameter is given by
$q_{EA}=D_{aa}(\tau\rightarrow\infty)$.

The problem defined by Eqs.(\ref{seff},\ref{scspin}) is still too complicated
to be solved exactly, even in the paramagnetic phase. However, it can be
related to a solvable problem, which can be shown to have the same qualitative
phase diagram and the same low-frequency and low-temperature universal
properties. To arrive at this solvable model, we have to go through two
additional steps. The first one is to "integrate out" conduction electrons
in $S_{eff}$, so that an action involving only spin degrees of freedom is
obtained. This cannot be done exactly because of the Kondo interaction,
but can be done asymptotically at low-energies by following the classic
Anderson-Yuval-Hamann approach to the Kondo problem \cite{AYH}. This approach
consists in separating the Kondo term into an Ising part $S^zs^z$ and a
spin-flip part, and performing an expansion in the spin-flip term to all
orders. The result of this procedure is a mapping of the Kondo part of
$S_{eff}$ onto an action involving an effective interaction for the
Ising components of the spins. This effective interaction is retarded and
decays as $1/\tau^2$ for large (imaginary time) separations $\tau$. Thus,
low-energy properties of the model can be studied by replacing $S_{eff}$ with:
\begin{equation}
S_{eff}' =
\int_0^{\beta} d\tau \int_0^{\beta} d\tau'\sum_{a,b}
S_a^z(\tau) [K(\tau-\tau') \delta_{ab} - J^2 D_{ab}(\tau-\tau')] S_b^z(\tau')
\label{seffprime}
\end{equation}
in which $K(\tau)\sim 1/\tau^2$ at large $\tau$,
{\it i.e} $K(i\omega_n)\sim \kappa\omega_n sgn(\omega_n)$ ($\kappa$ is a
dimensionless parameter depending on the anisotropy of the Kondo coupling).
This behaviour must be cutoff at high frequency $\omega > \Lambda$
({\it i.e} at short time separation). Because the decay
$K(\tau)\sim 1/\tau^2$ only holds for times up to $1/T_K$, where
$T_K$ is the `bare' Kondo temperature in the absence
of the exchange $J$, the cutoff must be chosen as:
$\Lambda\simeq T_K$. This slow decay of $K(\tau)$ stems from the
metallic nature of the spin-glass problem under consideration.
In the insulating case considered in \cite{MH,YSR}, the decay is
slower with $K(i\omega_n)\sim \omega_n^2$.

\section{Solution and physical properties}

\subsection{Solvable $M=\infty$ limit}

In order to analyze the effective action (\ref{seffprime}), we can follow
Ye, Sachdev and Read \cite{YSR} and generalize the model from Ising spins to
$M$-components quantum rotors $\widehat{n}_{i\mu}$, $\mu=1,\cdots,M$,
with $\widehat{n}_i^2=1$ at each site. The Ising case formally corresponds
to $M=1$, but the $M\rightarrow\infty$ limit will be first solved
exactly. Enforcing the local constraint by
Lagrange multipliers $\lambda_j$, the model is solved in this limit
by a saddle-point method \cite{YSR}, with
$i\lambda_j\equiv \lambda$ uniform at the saddle point. The free energy of
the model reads, in the paramagnetic (replica symmetric) phase:
\begin{equation}
{{F}\over{NM}} = \lambda + {{1}\over{2\beta}}\sum_n
\mbox{ln}\,[\lambda+K(i\omega_n)-
J^2 D(i\omega_n)]
\label{eqf}
\end{equation}
where the spin-correlator $D(i\omega_n)$ is given by:
\begin{equation}
D(i\omega_n) = {{1}\over{2J^2}} \{\lambda+K(i\omega_n) -
\sqrt{(\lambda+K(i\omega_n))^2-4J^2}\}
\label{eqd}
\end{equation}
The Lagrange multiplier $\lambda$ is determined by the constraint equation
(equivalent to setting $\partial F/\partial\lambda=0$):
\begin{equation}
-{{1}\over{2\beta}}\sum_n D(i\omega_n) =1
\label{eqlamb}
\end{equation}

\subsection{Connection with $z=2, d=3$ quantum critical phenomena}

In order to analyze the phase diagram and critical behaviour resulting from
these equations, it is very useful to put them in a different form, which
will reveal a connection with a different problem already analysed in
the literature. Indeed, instead of using a mapping onto the single-site
action (\ref{seffprime}), we could have solved the $M=\infty$
rotor model directly
for the fully-connected lattice. In this approach, the spin correlation
function $D(i\omega_n)$ is given by the on-site component of the
inverse of the random matrix: $(\lambda+K(i\omega_n))\delta_{ij}-J_{ij}$.
In the thermodynamic limit $N=\infty$, the eigenvalues of the matrix $J_{ij}$
have a semi-circular distribution given by:
\begin{equation}
\rho_J(x) = {{1}\over{2\pi J^2}}\,\sqrt{4J^2-x^2}\,
\theta(2J-|x|)
\label{scdos}
\end{equation}
Hence, the free-energy and spin propagator can be written in the
equivalent form:
\begin{eqnarray}
{{F}\over{MN}} = \lambda + \sum_n \int_{-2J}^{+2J} dx \rho_J(x) \,
\mbox{ln} [\lambda+K(i\omega_n)-x]\\
D(i\omega_n) = \int_{-2J}^{+2J} dx {{\rho_J(x)}\over{\lambda+K(i\omega_n)-x}}
\end{eqnarray}
Independently of the low-frequency behaviour of $K(i\omega_n)$ (metallic
or insulating), it is clear that only the solutions with $\lambda\geq 2J$
are admissible \cite{YSR} (so that no pole is encountered in the integration
over $x$). The critical boundary with the spin-glass phase is signalled
by $\lambda(T,J)$ reaching the value $2J$ (the condition $\lambda(T,J)=2J$
holds throughout the spin-glass phase). Thus, the combination
$\Delta\equiv\lambda-2J$ plays the role
of the important low-energy scale (which vanishes in the spin-glass phase).
Universal low-frequency properties are found when this energy scale is small.
These universal properties only depend on the low-frequency behaviour
of $K(i\omega_n)$ and on the fact that the
spectral density $\rho_J(x)$ has a square-root behaviour near its upper
band edge $+2J$. Formally, this behaviour is identical to that of fermions
with a kinetic energy proportional to $k^2$ in $d=3$ dimensions: we could
as well set $K(i\omega_n)+\lambda-x\equiv K(i\omega_n)+\Delta+k^2$ and
replace the integration over $x$ in the above equations by an integration over
$d^3k$ (with some upper cutoff). In this analogy, the scaling of frequency
with respect to $k$ is obtained from $K(i\omega_n)\sim \omega_n^2 \sim k^2$
in the insulating case of Ref.\cite{YSR}, while
$K(i\omega_n)\sim |\omega_n| \sim k^2$ in the metallic case.
Therefore, we conclude that there is a formal equivalence  with
quantum critical phenomena \cite{JH,AM} in the $d=3, z=1$ universality class
for the insulating case, and in the $d=3, z=2$ universality class for
the metallic case of interest here. The insulating case considered
in \cite{YSR} corresponds to a (quantum) Landau-Ginzburg model at its upper
critical dimension $d+z=4$ (with $\phi^4$ a marginal perturbation), while
the metallic case amounts to look at this model above its upper critical
dimension (with $\phi^4$ a dangerously irrelevant perturbation).
For these reasons, many results that will be derived below
for the metallic case are formally identical to
those of Millis \cite{AM} for the $d=3,z=2$ case.

In order to derive these
results, it is most useful to convert the Matsubara sums in the above
equations into real-frequency integrations. Using
$K(\omega+i0^+)\sim i\omega\, sgn(\omega)$ (with an upper cutoff $\Lambda$),
we can write the free energy as:
\begin{equation}
{{F}\over{NM}}\,=\lambda + \int_0^{\Lambda} {{d\omega}\over{\pi}}
\mbox{coth} {{\beta\omega}\over{2}} \int_{-2J}^{+2J} dx \rho_J(x)
\,\mbox{tan}^{-1} {{\omega}\over{\lambda-x}}
\label{eqfreal}
\end{equation}
and the constraint equation reads:
\begin{equation}
\int_0^{\Lambda} d\omega\, \chi''(\omega)\, \mbox{coth}
{{\beta\omega}\over{2}} \,=\,1
\label{eqscreal}
\end{equation}
with:
\begin{equation}
\chi''(\omega) \equiv -{{1}\over{\pi}} \mbox{Im} D(\omega+i0^+) =
sgn(\omega)\,\int_{-2J}^{+2J} dx
\rho_J(x) {{\omega}\over{\pi[\omega^2+(\lambda-x)^2]}}
\label{eqchireal}
\end{equation}

\subsection{Phase diagram}

\begin{figure}
\caption{Schematic phase diagram at mean-field level, as a function of
$T/J$ and $J_c/J$ (with $J_c\simeq\Lambda\simeq T_K$). The plain line is
the critical boundary with the spin-glass phase. All other lines are crossover
lines, corresponding to the regimes described in the text. The hatched region
is that of classical behaviour near the critical boundary.}
\label{phasediag}
\end{figure}
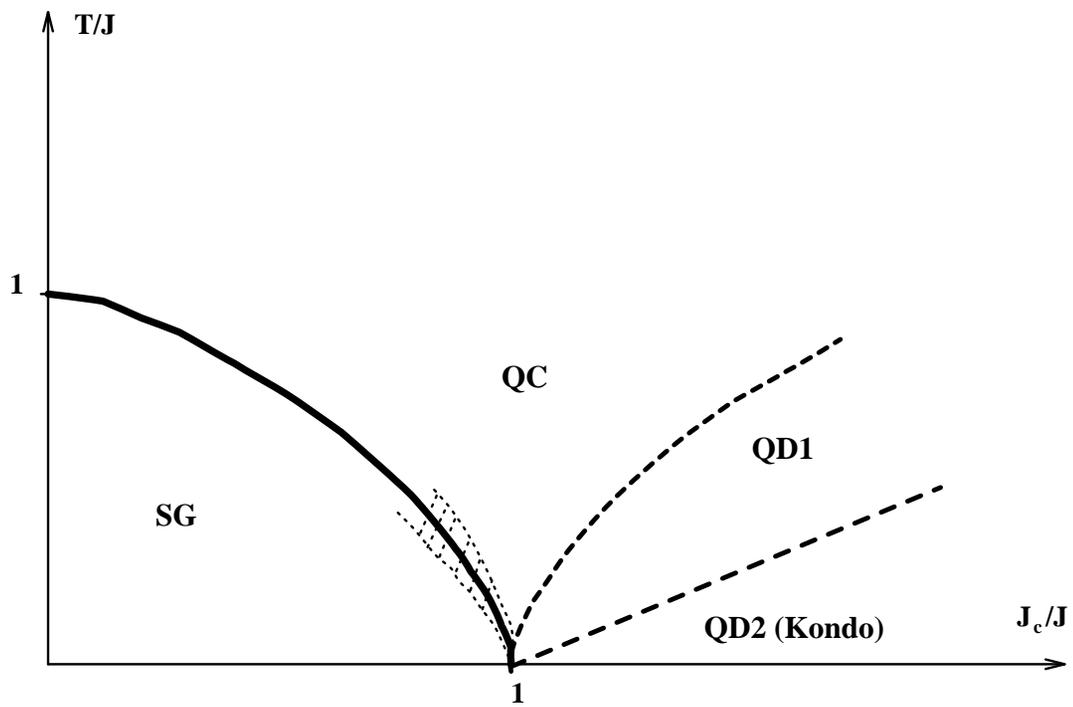

The phase diagram resulting from Eqs.(\ref{eqscreal},\ref{eqchireal})
is depicted schematically in Fig.\ref{phasediag}. It is qualitatively
similar to the insulating case \cite{YSR},
but the equations for the critical boundary and the various
crossover lines are affected by the different low-frequency behaviour
of $K(i\omega_n)$. Some calculations are detailed in the Appendix.
For very large $J$, the spin-glass transition
temperature is at $T_c=J$, the classical value \cite{SK}. $T_c$
decreases upon increasing quantum fluctuations ({\it i.e} increasing
$J_K$) and eventually vanishes for $J$ smaller than a critical
value: $J_c \simeq\Lambda \kappa^{1/3} \simeq T_K \kappa^{1/3}$.
Near zero temperature, the phase boundary is such that:
$1-J_c(T)/J_c \sim (T/\Lambda)^{3/2} \sim (T/T_K)^{3/2}$
(to be contrasted with $\sim T^2$ in the insulating case \cite{YSR}).

Next, we discuss the various crossover regimes \cite{AM} near the $T=0$ quantum
critical point at $J=J_c$ ({\it cf.} Fig.\ref{phasediag}).
Close to this point,
and for low frequency and small $\Delta$ (but $\omega/\Delta$ arbitrary),
the imaginary part of the local dynamical susceptibility $\chi''(\omega) =
-1/\pi Im D(\omega+i0^+)$ takes the scaling form:
\begin{equation}
\chi''(\omega) = {{C}\over{J_c^{3/2}}}\,sgn(\omega) \sqrt{|\omega|}\, f\left(
{{\omega}\over{\Delta}}\right)
\label{scal}
\end{equation}
The scaling function $f(x)$ is easily obtained from (\ref{eqd}) as:
\begin{equation}
f(x)=\,\left(x/(1+\sqrt{1+x^2})\right)^{1/2}
\end{equation}
and behaves as $f(x)\sim 1$ for
$x\rightarrow\infty$ and $f(x)\sim\sqrt{x}$ for $x\rightarrow 0$.
Hence, spin excitations are gapless in all regimes, with
$\chi''(\omega)\sim \omega/(J_c^{3/2}\sqrt{\Delta})$ for $\omega<<\Delta$ and
$\chi''(\omega)\sim sgn(\omega) \sqrt{\omega}/J_c^{3/2}$ for $\omega>>\Delta$.
The former, linear behaviour is characteristic  of local spin correlations
in a Fermi liquid with a low-energy scale $\Delta$, while the latter (which
holds down to $\omega=0$ at the $T=0$ critical point)
deviates strongly from Fermi-liquid theory.
$C$ is a non-universal constant, and the low-energy scale
$\Delta(T,J)\,\equiv\lambda-2J$ has different behaviour
in different regions of the phase diagram (Fig.\ref{phasediag}).

$\Delta=0$ inside the spin-glass phase, in which
$\chi''(\omega)\sim sgn(\omega) \sqrt{\omega}$ with no characteristic scale.

Raising temperature at $J=J_c$, one
enters the {\it quantum critical} (QC) regime, in which the physics
is dominated by the $T=0$ quantum critical point.
The energy scale $\Delta\equiv \lambda-2J$ is set entirely
by temperature in this regime and is found to be
$\Delta\sim T^{3/2}/\sqrt{J_c}\sim T^{3/2}/\sqrt{T_K}$. One would have naively
expected $\Delta\sim T$ but we find a {\it violation of this scaling} in
the QC regime of this model. Following the connection explained above,
this is an effect of the $\phi^4$ term in the quantum Landau-Ginzburg
model being a dangerously irrelevant perturbation for $d+z=5 > 4$.

Decreasing $J$ (or increasing $T_K$) from the
$T=0$ critical point, at low enough temperature, one enters the {\it quantum
disordered} (QD) regime in which $\Delta\sim J_c-J\sim T_K-J$ (to dominant
order). The crossover between the QC and QD regimes occurs at $J=J^*(T)$
obtained by matching the two behaviour of $\Delta$ given above,
with the result: $1-J^*(T)/J_c \sim (T/J_c)^{3/2} \sim (T/T_K)^{3/2}$.
As shown below, two distinct regions corresponding to $T\gg\Delta$ and
$T\ll\Delta$ must actually be distinguished within the QD regime, in which
the physical quantities have quite different low-temperature behaviour.
In the low-temperature region of the QD regime (denoted $\mbox{QD}_2$ on
Fig.\ref{phasediag}), the physics is that of a metal showing
Fermi-liquid behaviour below the coherence scale (or {\it effective}
Kondo scale) $\Delta$, which can be very small because of the
competition between the Kondo effect and the freezing of the local moments.
However, this interpretation has to be handled with some care, since this
scale enters the various physical quantities in quite different manners,
as detailed below.

Finally, near the phase boundary, there is a {\it classical} regime
\cite{AM} for $|J-J_c(T)|<<T^2/J_c$ dominated by purely classical
fluctuations and in which $\Delta\sim J_c(J-J_c(T))^2/T^2$.

\subsection{Specific heat, Susceptibility and NMR relaxation rate}

We now investigate the low-temperature behaviour of the specific heat
in the various regimes described above. Some details of the calculation
are provided in the Appendix, but the results could also be directly
read off from those in \cite{AM}, given the equivalence with the $d=3,
z=2$ problem. In the QC regime, the specific-heat coefficient
$\gamma\equiv C/T$ is found to behave as:
\begin{equation}
\gamma = \gamma_0 - {{A}\over{J_c^{3/2}}} \sqrt{T} + \cdots
\label{gamma}
\end{equation}
Interestingly, the non-Fermi liquid nature of the quantum critical
point results in a non-analytic correction to the low-temperature
specific heat in this regime.
Very close to the critical boundary with the spin-glass phase,
there is an additional contribution \cite{AM}:
$B T/[J-J_c(T)+T^{3/2}]^{1/2}$ ($\sim T^{1/4}$ for $J\simeq J_c(T)$),
which becomes rapidly negligible however as one moves away from the
critical boundary.

The non-analytic $\sqrt{T}$ behaviours continues
to hold within the QD regime as long as $T\gg \Delta$, {\it i.e}
$T\gg J_c-J$. This defines an additional subdivision of the QD
regime, corresponding to region $\mbox{QD}_1$ in Fig.\ref{phasediag}.
In the low-temperature ($\mbox{QD}_2$) region of the QD regime, (\ref{gamma})
is replaced by:
\begin{equation}
\gamma= \gamma_0 - {{C}\over{J_c^{3/2}}} \sqrt{\Delta}-
{{D}\over{J_c^{3/2}}} {{T^2}\over{\Delta^{3/2}}} +\cdots
\label{specqd}
\end{equation}
with $\Delta\sim J_c-J$ and $\gamma_0\propto 1/J_c$ remains finite as
$J\rightarrow J_c$.

Next, we discuss the low-temperature behaviour of the uniform spin
susceptibility $\chi$. This quantity is assumed to be measured by
applying a uniform field such that the Zeeman
energy $\mu H$ is smaller than the bare Kondo scale $T_K$. Then,
the Kondo effect still takes place
and the low-frequency behaviour of the resulting propagator $K(i\omega_n)$
is essentially unaffected. We can thus simply introduce
a uniform magnetic field $H\sum_i S_i^z$ in the effective
Ising model. The susceptibility is given by the sum over all pairs of sites
of the spin-spin correlation function taken at $\omega_n=0^+$, namely:
namely:
\begin{equation}
\chi\,=\,\sum_{ij}\overline{ [\lambda+K(i0^+)-J_{kl}]^{-1}_{ij} }
\end{equation}
In this expression, the overbar denotes an averaging over disorder.
The inverse of the random matrix present in this expression can be
evaluated by expanding in powers of $J_{kl}$. Because these couplings
have random signs and zero-mean, only closed paths (with $i=j$) give
a non-zero contribution on the fully-connected (or Bethe) lattice. Hence,
the uniform susceptibility behaves in an identical manner to the
{\it local} spin susceptibility:
$\chi_{loc} = \int_{0}^{\beta} d\tau D(\tau)$.
Note that this relies crucially on the fact that we are dealing with a
random system, and would not apply to a a uniform antiferromagnet for
example. Hence the formal analogy between the present problem and
the $z=2, d=3$ quantum antiferromagnet does not apply to the calculation
of the susceptibility. $\chi_{loc}$ (and thus $\chi$) is easily
obtained by setting $\omega_n \rightarrow 0^{+}$ in Eq.(\ref{eqd}) and
taking the real part. This yields
$2 J^2 \chi_{loc}= \mbox{const}. -2 (J\Delta(T,J))^{1/2} + \Delta(T,J)$.

In the QC regime, we must set $\Delta\sim T^{3/2}$, so that:
\begin{equation}
\chi(T) = \chi_0 (1- a ({{T}\over{J_c}})^{3/4} +\cdots )
\label{eqchit}
\end{equation}
with $\chi_0\sim 1/J_c$, and $c$ a numerical constant.  A non-analytic
correction departing from standard Fermi-liquid theory is again found.
In the high-temperature part of the QD regime ($\mbox{QD}_1$ region),
we have ({\it cf.} Appendix): $\Delta(T)=\Delta+T^{3/2}$, so that:
\begin{equation}
\chi(T) = \chi_0 - a_1{{\sqrt{\Delta}}\over{J_c^{3/2}}} -
a_2 {{T^{3/2}}\over{J_c^2\sqrt{\Delta}}} + \cdots
\label{eqchiqd1}
\end{equation}
In the low-temperature (Kondo) region of the QD regime, we have
({\it cf} Appendix): $\Delta(T)=\Delta + T^2/\sqrt{\Delta}$, so that:
\begin{equation}
\chi(T) = \chi_0 - a_1{{\sqrt{\Delta}}\over{J_c}} -
a_2 {{T^{2}}\over{J_c^2\Delta}} + \cdots
\label{eqchiqd2}
\end{equation}
In both (\ref{eqchiqd1}) and (\ref{eqchiqd2}), $\Delta\sim J_c-J$ and we
emphasize that $\chi_0\sim 1/J_c$ is non-singular as the critical point
is reached.

Finally, we consider the NMR relaxation rate, which is directly related
to the $\omega=0$ behaviour of the local dynamical susceptibility through:
\begin{equation}
{{1}\over{T_1T}} \equiv {{\chi''(\omega)}\over{\omega}}|_{\omega=0}
\end{equation}
Hence, this quantity always feels the linear regime $\chi''(\omega)\sim
\omega/\sqrt{\Delta}$ in the scaling form (\ref{scal}). In the
QC regime, this yields a temperature dependence of the relaxation rate
which differs from the usual Korringa behaviour ($1/T_1T\sim \mbox{const}.$)
found in a Fermi liquid:
\begin{equation}
{{1}\over{T_1T}} \sim {{1}\over{T^{3/4}}}
\label{nmrnfl}
\end{equation}
In contrast, in both the $\mbox{QD}_1$ and the $\mbox{QD}_2$ (`Kondo') regime,
the Korringa law is obeyed, but with an enhanced rate:
\begin{equation}
{{1}\over{T_1T}} \sim {{1}\over{\sqrt{\Delta}}} \sim
{{1}\over{\sqrt{J_c-J}}}
\end{equation}
We emphasize that in the same regime, the $T=0$ uniform spin susceptibility
is {\it not} critically enhanced (the low-energy scale $\Delta$ only enters
subleading corrections).

\subsection{Ising ($M=1$) case}

Before leaving model (\ref{ham1}), we would like to show that
the Ising case $M=1$ actually has the same universal
low-frequency behaviour
than the rotor model in the $M\rightarrow\infty$ limit that we have
analysed in detail. In order to show this, one possibility
is to adapt the reasoning of Refs.\cite{MH,YSR} to the present
case. Specifically, we can define the spin
irreducible self-energy associated with (\ref{seffprime}) by:
$D(i\omega_n)^{-1}=K(i\omega_n)-J^2 D -\Pi(i\omega_n)$.
For $M=\infty$, $\Pi(i\omega_n)$ reduces to a
constant $\Pi=\lambda$. To order $1/M$, the first contribution to
$Im \Pi$ is from decays into three spin waves. With
$D(\tau)\sim 1/\tau^{3/2}$ at the critical point, this leads to
$Im \Pi \sim \omega_n^{7/2}$. No non-analyticity with a weaker power
of frequency is induced to any order in $1/M$, and hence the low-frequency
behaviour found above at the critical point is unchanged. Another line
of reasoning leading to the same conclusion is to use the formal
equivalence with the $d=3, z=2$ universality class. Going from the
$M=\infty$ to the $M=1$ case just changes the specific coefficients
of the various terms of the (quantum) Landau-Ginzburg model, but
the equivalence holds for all $M$.

\section{Model with a single conduction electron fluid
and transport properties}
\label{mod2}

Finally, we shall release the assumption of independent Kondo
baths at each site, and consider a model with a single species
of conduction electrons. We shall consider a lattice of connectivity $z$,
and study the hamiltonian:
\begin{equation}
H_2\,=-\sum_{ij\sigma} t_{ij} c^+_{i\sigma}c_{j\sigma}
+ J_K \sum_i \vec{S}_i\cdot\vec{s}(i) -
\sum_{<ij>}J_{ij}S_i^z\cdot S_j^z
\label{ham2}
\end{equation}
The random exchange couplings are distributed according to a gaussian
distribution as above, which is now normalized such that
$\overline{J_{ij}^2}=J^2/z$. The limit of large lattice connectivity
$z\rightarrow\infty$ will be considered below. In this limit, the various
Green's functions satisfy self-consistent dynamical mean-field equations
\cite{GK,GKS}. These equations reduce the model to the solution of
a single-site effective action, which has again precisely the form in
Eq.\ref{seff}. However, there is now a self-consistency condition on
{\it both} $D(i\omega_n)$ (given by Eq.\ref{scspin} above) and
${\cal G}_0(i\omega_n)$, which is not known explicitly in contrast to
the case above. This self-consistency condition
depends on the specific lattice and
hopping term $t_{ij}$. More precisely, if $D(\epsilon)$ stands for the
non-interacting density of states of the lattice under consideration,
the effective propagator ${\cal G}_0$ must be such that the following
self-consistency equation holds \cite{GK,GKS}:
\begin{equation}
G_c(i\omega_n) = \int_{-\infty}^{+\infty} d\epsilon
{{D(\epsilon)}\over{i\omega_n+\mu-\Sigma_c(i\omega_n)-\epsilon}}
\label{scfull}
\end{equation}
In this equation, $G_c$ stands for the local conduction electron Green's
function $G_c\equiv -<T c c^+>_{Seff}$, and $\Sigma_c$ for the self-energy
$\Sigma_c\equiv {\cal G}_0^{-1}-G_c^{-1}$. Both should be viewed as
functionals of ${\cal G}_0(i\omega_n)$ (and $D(i\omega_n)$).

A first possibility  is to consider a model with a specific form of
{\it long-range} hopping (described in \cite{GKS}),
such that the non-interacting conduction electron
density of states is a Lorentzian $D(\epsilon)=1/\pi(\epsilon^2+\Gamma^2)$.
In this case, the Hilbert transform in (\ref{scfull}) yields:
$G_c = [i\omega_n+\mu-\Sigma_c+i\Gamma sgn(\omega_n)]^{-1}$, so that
$\Sigma_c$ disappears altogether from the self-consistency equation and
${\cal G}_0$ is actually known explicitly as before:
${\cal G}_0^{-1}=i\omega_n+i\Gamma sgn(\omega_n)$.
Hence, {\it exactly the same equations} as in
the above model (\ref{ham1}) with independent Kondo baths are found for
this model with long-range hopping, and no non-trivial feedback
of the conduction electron dynamics into the spin dynamics is possible.

The situation is different for a model with {\it short-range hopping}
of the conduction electrons. For definiteness (but without loss of
generality), we may consider the $z=\infty$ Bethe lattice, with
nearest-neighbour hopping normalised according to $t_{ij}=t/\sqrt{z}$.
(This corresponds to a semi-circular d.o.s with a half-width $2t$).
In this case, (\ref{scfull}) takes the simpler form:
\begin{equation}
{\cal G}_0^{-1}(i\omega_n)=i\omega_n+\mu-t^2 G_c(i\omega_n)
\label{scc}
\end{equation}
$G_c$ has to be determined from the solution of $S_{eff}$ itself, so that
the problem involves a self-consistency condition on {\it both}
$D(i\omega_n)$ and ${\cal G}_0(i\omega_n)$.
The main question is whether this "feedback" of the non-trivial spin
dynamics into ${\cal G}_0$ can change the low-frequency behaviour of
$\chi''(\omega)$ at the $T=0$ critical point. We shall give an argument
that this is not the case, and that (\ref{scal}) still holds.

Let us imagine that the coupled problem (\ref{seff},\ref{scspin},\ref{scc})
is solved iteratively, starting
from a ${\cal G}_0$ which has the same long-time behaviour than in the
model without feedback, namely ${\cal G}_0(\tau)\sim 1/\tau$. Then,
the arguments above yield $D(\tau)\sim 1/\tau^{3/2}$ at the critical
point. Inserting this into the effective action (\ref{seff}), we
have to compute the conduction electron Green's function $G_c$ and feed
it back into the self-consistency condition (\ref{scc}) to see how
${\cal G}_0$ is affected.
In order to find the behaviour of $G_c$, it is convenient to use a
representation of the localised spins
by pseudo-fermions $f_{\sigma}$, such that:
$\vec{S}=f^{+}_{\sigma}\vec{\sigma}_{\sigma\sigma'} f_{\sigma} /2$. This
amounts to "undoing" the Schrieffer-Wolff transformation and representing
the original Kondo lattice as a periodic Anderson model with a very
large $U$ in the local moment limit $\epsilon_f\simeq -U/2$.
We shall not attempt here to find the actual behaviour of
the $f$-electrons Green's functions $G_f(\tau)$, but
it is easily seen that it cannot decay faster than $1/\tau^{3/4}$.
Indeed, $D(\tau)$ always contains a term
of the form $G_f(\tau)G_f(-\tau)$ (supplemented by vertex corrections),
so that a decay slower than $1/\tau^{3/4}$ is inconsistent with
$D(\tau)\sim 1/\tau^{3/2}$. Furthermore, the slowest possible
decay of $G_f(\tau)$ is the Fermi liquid form $G_f(\tau)\sim 1/\tau$.
Hence, we conclude that $G_f(\tau)\sim 1/\tau^{\theta}$, with the
exponent $\theta$ such that $3/4\leq \theta \leq 1$.
Now, the t-matrix associated with
(\ref{seff}) is proportional to $G_f(i\omega_n)$, so that the conduction
electron Green's function is given by:
$G_c(i\omega_n)=
{\cal G}_0(i\omega_n)+ V^2{\cal G}_0(i\omega_n)^2 G_f(i\omega_n)$.
Hence $G_c(\omega)$ behaves as $\omega^{\theta-1}$ at low-frequency.
Inserting this into the self-consistency equation (\ref{scc}) in order to
see how ${\cal G}_0$ is affected at the next step of the iteration, we
see that ${\cal G}_0(\tau)$ behaves as $1/\tau^{2-\theta}$ for large
$\tau$. Because of the constraint above, this exponent satisfies
the bounds $1\leq 2-\theta \leq 1+1/4$, so that ${\cal G}_0$ cannot decay
faster than $1/\tau$ at the next iteration. Hence, the associated spectral
density $Im {\cal G}_0(\omega+i0^+)$ cannot diverge at $\omega=0$: it is
either finite or vanishes. For large enough Kondo coupling, this yields a
standard Kondo effect \cite{WF}, and hence an effective interaction
between Ising spins such that $K(\tau)\sim 1/\tau^2$ as before, so that the
behaviour of $\chi''(\omega)$ at the critical point remains unchanged
at any step of the iterative solution of the coupled equations.

A perturbative argument can actually be given that the dominant behaviour
$G_c(\tau)\sim 1/\tau$
of the conduction electron Green's function is not affected at the $T=0$
critical point, {\it i.e} that $\theta=1$. Indeed, if we treat the residual
coupling between the conduction electrons and localised spins in second-
order perturbation theory, we obtain a contribution to the self-energy:
$\Sigma_c(\tau)\propto J_K^2 D(\tau){\cal G}_0(\tau) \sim 1/\tau^{5/2}$,
so that $\Sigma_c(\omega)$ behaves as $\omega^{3/2}$ at $T=0$. Hence, this
scattering is unable to modify the dominant term in the long-time behaviour of
$G_c(\tau)\sim 1/\tau$.

It has however important consequences for the transport properties near
the $T=0$ critical point, as we shall now show. We first perform a
more precise evaluation of the finite-temperature scattering rate to
order $J_K^2$, which takes the form:
\begin{equation}
\mbox{Im}\Sigma_c(\omega+i0^+) \propto
J_K^2 \rho_0(0) \int_{-\infty}^{+\infty} du\,\chi''(u)
\left({{1}\over{e^{\beta u}-1}}+{{1}\over{e^{\beta(u+\omega)}+1}}\right)
\end{equation}
Inserting the scaling form (\ref{scal}), this leads to the following
low-frequency and low-temperature dependence, in the quantum critical
regime:
\begin{equation}
\mbox{Im}\Sigma_c(\omega+i0^+) \propto
{{J_K^2}\over{J_c^{3/2}}}\rho_0(0) [\omega^{3/2}+T^{3/2}]
\end{equation}
while in both regions of the QD regime:
\begin{equation}
\mbox{Im}\Sigma_c(\omega+i0^+) \propto
{{J_K^2}\over{J_c^{3/2}}}\rho_0(0) [{{\omega^2}\over{\sqrt{\Delta} }}+
{{T^2}\over{\sqrt{\Delta}}}]
\end{equation}
Since we are dealing with a model on a lattice with infinite connectivity,
the vertex corrections to the conductivity vanish \cite{AK}, and the
$dc$-conductivity is simply given by:
\begin{equation}
\sigma_{dc}\propto \int d\epsilon D(\epsilon) \int d\omega \,
{{Im\Sigma_c(\omega)}\over{(\omega+\mu-Re\Sigma_c-\epsilon)^2
+ (Im\Sigma_c)^2}}\,
{{\partial f}\over{\partial \omega}}
\end{equation}
Hence, the above calculation of the scattering rate leads to the following
non-Fermi liquid temperature dependence of the resistivity in the
QC regime :
\begin{equation}
\delta\rho(T) \sim T^{3/2}
\label{ronfl}
\end{equation}
while in both regions of the QD regime:
\begin{equation}
\delta\rho(T) \sim {{T^2}\over{\sqrt{\Delta}}}
\end{equation}
with $\Delta\sim J_c-J$. Hence a Fermi liquid behaviour $\delta\rho =A T^2$ is
recovered in the QD regime, but with a critically enhanced rate
$A\sim 1/\sqrt{\Delta}$. We emphasize that in the same regime, the specific
heat coefficient $\gamma$ is {\it not} critically enhanced. This
distinguishes the QD Fermi-liquid regime from conventional
heavy-fermion behaviour in which {\it both} $A$ and $\gamma$ are large for
small $T_K$, with \cite{CV} $A\propto\gamma^2$.

\section{Conclusion}

In this paper, we have studied Kondo lattice models with a quenched
random exchange between localised spins. The mean-field phase diagram
has been investigated (Fig.\ref{phasediag}) and found to display
several different regimes near the quantum critical point associated
with the $T=0$ spin-glass transition. In the `quantum critical' regime,
the specific heat coefficient and susceptibility display non-analytic
corrections to Fermi liquid behaviour given by (\ref{gamma},\ref{eqchit}),
while the NMR relaxation rate (\ref{nmrnfl}) and resistivity (\ref{ronfl})
have a non-Fermi liquid temperature dependence. In this regime,
the important low-energy scale violates naive scaling and varies as a
power of temperature ($\Delta \sim T^{3/2}$ at mean-field level).
In the low-temperature part of the quantum-disordered region (`Kondo regime'),
Fermi-liquid behaviour is recovered, but the NMR and scattering rate are
critically enhanced as the transition is reached, while $\gamma$ and $\chi$
are not.

These results may have qualitative relevance for $Y_{1-x}U_xPd_3$ and
related systems \cite{MBM,SML} since they indicate that non-Fermi liquid
behaviour is a rather generic feature associated with a $T=0$ spin-glass
transition in a metallic system. However, the reported experimental
behaviour ($\gamma \sim -\mbox{ln} T$, $\chi = \chi_0-\sqrt{T}$,
$\delta\rho\sim T$) is not in good agreement with our mean-field
results. This raises theoretical questions associated with the fluctuations
beyond mean field, and also experimental questions concerning the actual
investigation of the critical scaling regime.

\newpage
\acknowledgements
As this work was being completed, we became aware of a work by
S.Sachdev, N.Read and R.Oppermann \cite{SRO} in which the phase diagram
and crossovers described here are also analysed, and a detailed
theoretical investigation of the fluctuations beyond mean-field is performed.
A.S would like to acknowledge a discussion with S.Sachdev on these
topics, and the hospitality of Yale University on this occasion. We
would also like to thank D. Huse and A. Millis for useful discussions.

\appendix

\section{}

In this Appendix, we provide some details on the analysis of the various
crossovers and on the calculation of the specific heat. The starting point
are the Eqs.(\ref{eqfreal},\ref{eqscreal},\ref{eqchireal}), in which we
set: $\Delta \equiv \lambda-2 J$ and change variables in the integrations
over $x$ by setting $x=2J-\epsilon$.

We shall deal first with the constraint, Eq.(\ref{eqscreal}), which reads
(replacing $\rho_J(x)$ by its square-root form near the upper
band edge):
\begin{equation}
1\,=\,\int_{0}^{\Lambda} {{d\omega}\over{\pi}}
\int_{0}^{4J} {{d\epsilon}\over{\pi J^{3/2}}} \sqrt{\epsilon}\,
\mbox{coth}{{\beta\omega}\over{2}}\,
{{\omega}\over{\omega^2+(\epsilon+\Delta)^2}}
\end{equation}
The integral over $\omega$ needs to be cutoff at $\Lambda\sim T_K$,
while the integration over $\epsilon$ is ultra-violet convergent
and its upper limit could as well be set to $+\infty$.
We shall focus on the regions where $\Delta\ll T$ (which includes the QC
regime and the upper part of the QD regime). Under this assumption,
an expansion in $\Delta$ can be performed to yield:
\begin{equation}
1\,=\,\int_{0}^{\Lambda} {{d\omega}\over{\pi}}
\int_{0}^{4J} {{d\epsilon}\over{\pi J^{3/2}}} \sqrt{\epsilon}\,
\mbox{coth}{{\beta\omega}\over{2}}\,
\left(
{{\omega}\over{\omega^2+\epsilon^2}} -
\Delta \, {{2\omega\epsilon}\over{(\omega^2+\epsilon^2)^2}} \,
+O(\Delta^2) \right)
\end{equation}
In the integrations over $\epsilon$, we set $\epsilon=u\omega$ so that the
$\omega$ dependance becomes apparent. Then, we use
$\mbox{coth} \beta\omega/2 = 1+ 2/(e^{\omega/T}-1)$ and expand the
integrals involving the last term at low temperature.
The constraint equation then takes the form:
\begin{equation}
1\,=\,
A_1 ({{\Lambda}\over{J}})^{3/2} + A_2 ({{T}\over{J}})^{3/2}
- {{\Delta}\over{J}}\,
\left(B_1({{\Lambda}\over{J}})^{1/2}+B_2 ({{T}\over{J}})^{1/2}\right) \,
+O(\Delta^2)
\end{equation}
In this expression, the $A_i$'s and $B_i$'s are purely numerical,
positive constants (independant of the cutoff and of $J$).

The location of the quantum critical point is readily obtained by
setting $\Delta=T=0$, yielding: $J_c=A_1^{2/3}\,\Lambda \propto T_K$.
When $T$ is increased above this point, the behaviour of $\Delta$
in the QC regime
is found by cancelling the dominant terms to next order,
leading to: $\Delta \sim  A_2/B_1 T^{3/2}/\sqrt{\Lambda}$. The above
expansion is valid as long as $\Delta\ll T$, so that we can also
use it for $T^{3/2}\ll \Delta \ll T$, corresponding to the upper part
of the $QD$ region. In this case, we have to
expand in $J=J_c-\delta J$ and the behaviour
$\Delta \sim \delta J + O(T^{3/2})$ is obtained.
Also, the shape of the critical boundary at low temperature is found
by expanding in $\delta J = J_c(T)-J_c$, with $\Delta$ set to $0$.
This yields $\delta J \propto T^{3/2}/\sqrt{J_c}$.

Next, we give some indications on the low-temperature expansion of
the free-energy in the QC regime. We rewrite Eq.(\ref{eqfreal}) for the
free-energy per site $f=F/NM$ under the form:
\begin{equation}
f\,=\, \mbox{const.}\,+\,
\,\int_{0}^{\Lambda} {{d\omega}\over{\pi}}\,
\mbox{coth}{{\beta\omega}\over{2}}\,
\int_{0}^{4J} {{d\epsilon}\over{\pi J^{3/2}}} \sqrt{\epsilon}\,
\mbox{tan}^{-1} {{\omega}\over{\epsilon+\Delta}}
\end{equation}
In the QC regime, we should use the following scaling variables:
$\omega=T\widetilde{\omega}, \epsilon=T\widetilde{\epsilon},
\Delta=T^{3/2}\widetilde{\Delta}$. Thus it is clear that $\Delta$ is
the smallest energy scale, and we can simply expand the above expression
in powers of $\Delta$. The important point here is that {\it the linear
term vanishes} because of the constraint equation above. Hence,
this expansion reads:
\begin{equation}
f\,=\, \mbox{const.}\,+\,
\,\int_{0}^{\Lambda} {{d\omega}\over{\pi}}\,
\mbox{coth}{{\beta\omega}\over{2}}\,
\int_{0}^{4J} {{d\epsilon}\over{\pi J^{3/2}}} \sqrt{\epsilon}\,
\left(\mbox{tan}^{-1} {{\omega}\over{\epsilon}}\,+
\Delta^2\,{{\omega\epsilon}\over{(\omega^2+\epsilon^2)^2}}\,
+O(\Delta^3)\right)
\end{equation}
The coefficient of the $\Delta^2$ term is the same as found above in the
analysis of the constraint. It leads to a dependence of the form
$\Delta^2 (\mbox{const.}+\sqrt{T})\sim T^3 + T^{5/2}$.

In order to find the temperature dependence of the first term, we
make a low frequency expansion:
\begin{equation}
\int_{0}^{4J} {{d\epsilon}\over{\pi J^{3/2}}} \sqrt{\epsilon}\,
\mbox{tan}^{-1} {{\omega}\over{\epsilon}}\,\sim\,
{{\omega}\over{J}} +
({{\omega}\over{J}})^{3/2} + O(\omega^3)
\end{equation}
Hence, this term yields a contribution of order $T^2 + T^{5/2}$
to the free-energy. Overall, we find the low-temperature expansion:
\begin{equation}
f\sim \mbox{const.} + f_1 T^2 + f_2 T^{5/2} +\cdots
\end{equation}
in which the coefficients $f_i$'s are non-singular and non-vanishing
as the critical point $J=J_c$ is reached.
The behaviour of $C=T \partial^2 f/\partial T^2 \sim \gamma_0 T -T^{3/2}$
follows.

Finally, we comment on the behaviour of $\Delta$ in
the low-temperature $\mbox{QD}_2$ regime. In this regime, the above expansions
are no longer valid
since they assumed $T\gg\Delta$. Concentrating {\it e.g} on the
constraint equation, we see that the leading low-temperature correction
is now controlled by the long-time behaviour
$D(\tau)\sim 1/\tau^2\sqrt{\Delta}$, which holds for $\tau\Delta\gg 1$.
Using the Poisson summation formula, this yields:
\begin{equation}
{{1}\over{\beta}}\sum_n D(i\omega_n) |_{J=J_c,T=0} -
{{1}\over{\beta}}\sum_n D(i\omega_n) = a\delta J + b\delta\lambda
+c {{T^2}\over{\sqrt{\delta J}}} +\cdots
\end{equation}
with $\delta J=J-J_c$, $\delta\lambda=\lambda-2J_c$. Cancelling the
leading corrections, one obtains:
$\Delta= J_c-J +{{T^2}\over{\sqrt{J_c-J}}}$.

\end{document}